\documentstyle[11pt]{article}
\begin{document}

\begin{center}

{\huge PHYSICS AND COSMOLOGY} \\

{\huge IN AN INHOMOGENEOUS UNIVERSE} \\

({\it Contribution to the proceedings of the 49th Yamada conference} \\

{\it on black holes and high-energy astrophysics} \\

{\it to be published by Universal Academy Press, Tokyo}) \\

\bigskip

Andrzej KRASI\'NSKI \\ 

N. Copernicus Astronomical Center and College of Science \\

Polish Academy of Sciences, Bartycka 18, 00 716 Warszawa, Poland \\

email: akr@camk.edu.pl
 
\end {center}

\bigskip

\noindent {\bf Abstract}

Examples are presented of applications of the Lema\^{i}tre - Tolman model to problems of astrophysics and gravitation theory. 
They are: 1. Inferring the spatial distribution of matter by interpretation of observations; 
2. Interaction of inhomogeneities in matter distribution with the CMB radiation; 
3. Evolution of voids; 4. Singularities; 5. Influence of electromagnetic field on gravitational collapse. 
This review is meant to demonstrate that the theory is already well-prepared to meet the challenges posed by an inhomogeneous Universe.

\bigskip

\noindent {\bf 1. Why consider inhomogeneous models?}

The more appropriate question here is: why not? 
We know that the Friedmann - Lema\^{i}tre - Robertson - Walker (FLRW) models are a realistic first approximation to the geometry and physics of our Universe. 
We know (from many published examples, and see also below) that their generalizations are still reasonably simple. 
The urge to explore the unknown (once upon a time considered a necessary part of a scientist's personality) still lingers in the physics community. 
This should be a sufficient motivation for the beginning; more of it will appear along the way.

In this article, only one class of models will be discussed, the Lema\^{i}tre (1933 and 1997) - Tolman (1934 and 1997) (LT) models, and only a small selection of existing material will be presented. 
For a more complete treatment see the other review by this author (Krasi\'nski 1997).

\bigskip

\noindent {\bf 2. The Lema\^{i}tre - Tolman model}

Assume that the Universe is isotropic around one observer, but not spatially homogeneous. 
(How to save the "cosmological principle" in such a model will be explained the end of this section.)
Assume that the matter of the Universe is dust. 
In comoving coordinates, the metric form is then:

\begin{equation}
ds^2 = dt^2 - S^2(t,r)dr^2 - R^2(t,r)(d\vartheta ^2 + \sin ^2\vartheta d\varphi ^2). 
\end{equation}

\noindent The case $R,_{r} = 0$ has to be considered separately and it leads to the solution found by Datt (1938, see also Krasi\'nski 1997). 
It is an inhomogeneous generalization of the Kantowski - Sachs (1966) solution and has interesting geometrical properties, but has not been exploited for astrophysics, so it will not be discussed here. 
Assume now $R,_{r} \neq 0$. The Einstein equations then imply:

$$
S = R,_{r}/[1 + 2E(r)]^{1/2},
$$
$$
{R,_t}^2 = 2E(r) + 2M(r)/R + {1\over 3} \Lambda R^{2},
$$
\begin{equation}
8\pi G\rho /c^{2} = 2M,_{r}/(R^{2}R,_{r}),
\end{equation}

\noindent where $E(r)$ and $M(r)$ are arbitrary functions, $\Lambda $ is the cosmological constant and $\rho $ is the mass-density of dust. 
All formulae are covariant with the coordinate transformations $r = f(r')$, where $f$ is an arbitrary function.
Therefore it may be (and usually is) assumed that $r = 0$ at the center of symmetry. 
The equation defining $R$ can be solved in terms of the Weierstrass elliptic function (this was done by Lema\^{i}tre
already in 1933), and has the same algebraic form as in the Friedmann models. 
Indeed, the Friedmann models are contained in (2) as the following subcase:

\begin{equation}
E = {1\over 2}kr^{2},\qquad M = M_{0}r^{3},\qquad R = r{\cal R}(t), 
\end{equation}

\noindent where $k$ is the spatial curvature index, $M_{0}$ is the Friedmann mass integral and ${\cal R}(t)$ is the Friedmann scale factor. 
However, since $R$ is a function of both $t$ and $r$, a solution of the second equation in (2) will depend on $(t -
t_{B}(r))$, where $t_{B}(r)$ is one more arbitrary function (it is a constant in the Friedmann limit). 

In the LT model every shell of constant $r$ evolves
independently of the other ones, with its own initial conditions, but according to the same evolution law as in a Friedmann model. 
The function $M(r)$ specifies the initial distribution of mass, $E(r)$ specifies the initial distribution of energy, and $t_{B}(r)$ specifies the instant of the Big Bang, which is seen to be position-dependent.

The coordinate-independent definition of the Friedmann limit is:

\begin{equation}
(E^{3}/M^{2}),_{r} = 0 = t_{B,r} , 
\end{equation}

\noindent and the functions present in eq. (4) define the two types of perturbation of the Friedmann model that are well-known from studies of the linearized Einstein equations: 
the one in which $(E^{3}/M^{2}),_{r} \neq 0 = t_{B,r}$ is an increasing density perturbation, and the one in which $(E^{3}/M^{2}),_{r} = 0 \neq t_{B,r}$ is a decreasing density perturbation. 
(Increasing and decreasing refer to the solution in which matter expands away from the Big Bang.)

Let us now deal with the problem of the "cosmological principle". 
Since the Friedmann models are subcases of the LT model, the functions $E(r), M(r)$ and $t_{B}(r)$ can be chosen so that for $r \ge r_{0}$ they go over (differentiably to a desired order) into their Friedmann forms. 
In this way, the LT model becomes an isolated island in a Friedmann background, and the background does not feel the presence of the island if the matching hypersurface $r =
r_{0}$ is comoving. 
Hence, an arbitrary number of such islands in the same
background can be considered, and such a Universe model will be spatially homogeneous in a large scale while being inhomogeneous in the small scale. 
Hence, the cosmological principle does not prohibit the LT model. 
However, this principle is a philosophical construct and it should not bother anyone in exploring theoretical possibilities (see also Ellis 1984).

This section is a very short exposition which was meant to provide a minimal background for reading the remainig part of this article. 
The number of papers published on the LT models well exceeds 100, they are reviewed more extensively by this author elsewhere (Krasi\'nski 1997).

\bigskip

\noindent {\bf 3. Do we really know the distribution of matter in our Universe?}

The main part of this section are results of the paper by Kurki-Suonio and Liang (1992) (abbreviated KSL).

Let us assume that the LT model is used to interpret the astronomical observations. 
Observers receive their information about matter-distribution in the Universe along the light-cones. 
What they directly measure is the mass-density in the redshift space:

\begin{equation}
\hat{\rho }(z) = {H_0}^3 {dm\over dV(z)} = {H_0}^3 {1\over 4\pi z^{2}} {dm\over dz} ,
\end{equation}

\noindent where $H_{0}$ is the Hubble parameter at the observer's position at the time of observation (see below), $z$ is the redshift, $dV(z) = 4\pi z^{2}dz$ is the element
of volume of the redshift space and $dm$ is the amount of mass in $dV$. 
(The coefficient $H^{3}_{0}$ is there only for conversion to convenient units. The quantity $\hat{\rho }$ would be proportional to the physical mass-density if the space were flat and the Hubble law would exactly apply.) 
In order to infer the physical mass-density along the past light-cone, $\rho (z)$, and the mass-density at the present time, $\rho (t_{0},r)$, two quantities have to be read out from $\hat{\rho }$: the distance to the light-source, $l(z)$, and its velocity, $v(z)$. 
These cannot be measured independently, and so the problem is underdetermined. 
Without knowing $l$ and $v$ separately, we do not know the initial data for $R(t,r)$, and so the projection from the observer's past light-cone to the hypersurface $t = t_{0}$ is undetermined, too. 
The "knowledge" of these parameters in the Friedmann
models is spurious - deduced from pre-assumed symmetries.

Let us see what the LT model implies for this problem. 
For simplicity, it will be assumed that $\Lambda = 0$ and that the model is a perturbation of the $k < 0$ Friedmann model with the density parameter $\Omega _{0} = 0.1$. 
Then $E > 0$ and the solution of eq. (2) can be represented parametrically as:

\begin{equation}
R(t,r) = {M\over 2E} (\cosh \eta - 1), \qquad \sinh \eta - \eta = {(2E)^{3/2}\over M} [t - t_{B}(r)].
\end{equation}

\noindent If the observer is located at $r = 0$ and the light-source at $r = r_{\rm em}$, then the redshift and the Hubble parameter at the observer's position are given by:

\begin{equation}
\ln (1 + z) = \int_0^{r_{\rm em}} R,_{tr}[1 + 2E(r)]^{-1/2}dr,\qquad H_{0} = R,_{t}/R \hspace{1mm} _{\rule{0.1mm}{4mm} \hspace{1mm}(t,r)=(t_{0},r)}
\end{equation}

\noindent and $\hat \rho$ is related to the physical mass-density $\rho $ by:

\begin{equation}
\hat{\rho }(z) = {H_0}^3 {1\over 4\pi z^{2}} {dV\over dz} \rho (t,r) = {{H_0}^3 c^{2}\over 8\pi G} {2M,_{r}\over (1 + z)z^{2}R,_{tr}} ,
\end{equation}

\noindent where $dV = [4\pi R^{2}R,_{r}/(1 + 2E)^{1/2}]dr$ is the element of volume in the LT model. 
What can be said about the connection between $\hat{\rho }, \rho (z)$ and $\rho (t_{0},r)$ on the basis of information from observations?

For the beginning, let us choose the increasing perturbation with the arbitrary functions chosen so that the function $\hat{\rho }(z)$ has maxima at $z = 0.1$ and $z = 0.2$ 
of half-width $\Delta z = 0.02$ and amplitude $\Delta \hat{\rho }/\hat{\rho }_{\hom } = 16$ (see figures in KSL 1992), and elsewhere $\hat{\rho }/\hat{\rho }_{\hom } = 1$ ($\hat{\rho }_{\hom }$ is $\hat{\rho }$ calculated for the Friedmann background). 
In this case, the physical mass-density along the light-cone has maxima of amplitude $\Delta \hat{\rho}/\hat{\rho }_{\hom } = 5.2,$ i.e. observation exaggerates reality by a factor of more than 3. 
Moreover, the mass-density at the hypersurface $t = t_{0}$ has here maxima of unequal shape and amplitude: 
since the perturbation is increasing, and the farther condensation has had more time to evolve between the light-cone and the hypersurface $t = t_{0}$, the farther condensation has a greater amplitude and smaller half-width at $t = t_{0}$.

As a second example, let us take a decreasing perturbation with maxima of $\hat{\rho }(z)$ at the same redshifts and the same half-width as before, but with an amplitude of $\Delta \hat{\rho }/\hat{\rho }_{\hom } = 0.025$. 
This corresponds to {\it minima} in the physical $\rho (z)$ of amplitude $-0.37$, and to unequal minima in $\rho (t_{0},r)$, the farther one being broader and shallower. 
In this case, observation plainly contradicts reality.

Since the two kinds of perturbation influence the relation between $\hat{\rho }(z)$ and $\rho (z)$ in opposite ways, they can be tuned so that all inhomogeneity in $\hat{\rho }(z)$ is cancelled in $\rho (z)$. 
(The inhomogeneities in $\hat{\rho }(z)$ are in this case caused exclusively by "peculiar motions", not by inhomogeneities in the gravitational field.) 
Every such tuning will be unstable because each perturbation evolves according to a different law, and so the cancelling occurs only at a single light-cone; 
it does not survive to the hypersurface $t = t_{0}$, but the distribution of $\rho (t_{0},r)$ is different from $\hat{\rho }(z)$. 
An interesting example is given in the KSL 1992 paper: the function $\hat{\rho }(z)$ was chosen identical to the density as a function of redshift in the direction
of the North Galactic Pole from the Deep Redshift Survey 
(Broadhurst et al. 1990), and the functions $M(r), E(r)$ and $t_{B}(r)$ could still be chosen so that $\rho (z)$ was the same as in the unperturbed Friedmann model.

The KSL paper 1992 concludes with a statement that the present author finds to be the most important memento of this whole review: "$\ldots $it is fundamentally self-inconsistent to use a Friedmann relation between redshift and comoving distance when dealing with inhomogeneities. 
$(\ldots )$ What we see on the past light cone are only momentary "mirages"". 
This inconsistency was felt by several authors already in the 1930ies (see Krasi\'nski 1997, the list would include Lema\^{i}tre, Tolman, McVittie, Dingle, W. H. McCrea and still a few more). 
The fundamentalist faith in the "cosmological principle" is of later origin.

Considerations of similar kind were published by Moffat and Tatarski (1992 and 1995) and by Ribeiro (1992 and 1993). 
In particular, Ribeiro showed that the function $\rho (l)$ (i.e. density vs. luminosity distance) calculated for any Friedmann model contradicts what astronomers expect: it is nearly constant for small distances, and decreasing for large distances. 
It is not constant because the light-cone cuts through hypersurfaces of different mass-densities. 
It is decreasing (and in fact $\lim_{l \to \infty} \rho (l) = 0)$ because $l \to \infty $ when $z \to \infty $,
but the amount of mass within the $z = \infty $ hypersurface is finite. 
The conclusion from this is (at A. K.'s responsibility) that if the Universe looks homogeneous to astronomers, then this is an observational evidence against the Friedmann models.

\bigskip

\noindent {\bf 4. Anisotropies in the cosmic microwave background radiation generated by density inhomogeneities in the Universe}

The LT model is a convenient tool for considering this problem. 
The inhomogeneity is assumed to be at and around the center 
of symmetry (assumed to be at $r = 0$).
It may be a condensation or rarefaction, 
in both cases of arbitrary amplitude and density profile. 
As $r$ tends to infinity, the functions of the LT 
model tend to their Friedmann values. 
Light-rays are assumed to be emitted at the time of last scattering 
and received at the present time, 
but they pass through the inhomogeneity 
along different paths (i.e. at different 
distances from the center of symmetry). 
The question to which an answer is sought here is: 
how does the inhomogeneity influence the temperature 
of the radiation received by the observer? 
The answer is found by numerical intergration of the 
equations of null geodesics in the LT model. 
When the tangent vector to a null geodesic is already 
calculated, the temperature at the observation event is found from:

\begin{equation}
{{T_{\rm obs}}\over {T_{\rm em}}} 
= {1\over {1 + z}} = {{{(k^{\mu}u_{\mu})}_{\rm obs}} 
\over {{(k^{\mu}u_{\mu})}_{\rm em}}} ,
\end{equation}

\noindent where "obs" refers to the observation event, 
"em" refers to the emission event, $k^{\mu }$ is the 
tangent vector to the null geodesic and $u_{\mu }$ is 
the velocity vector of the dust medium; 
$T_{\rm em}$ is the same for all rays.

This method of investigation was first 
proposed by Raine and Thomas (1981) and first applied with a 
conclusive result by Panek (1992, only for the $k = 0$ 
Friedmann background). The results quoted below were obtained in the course 
of an elaborate project by Arnau et al. (1993 and 1994), 
S\'aez et al. (1994) and Fullana et al. (1996). 
The density amplitudes, profiles and sizes of the inhomogeneities 
were chosen so as to correspond to their current values in the observed 
objects (voids, galaxy clusters, the Great Attractor), 
but the $\Omega $-parameters of the background Friedmann 
regions were varied from one calculation to another.
Any condensation in the Universe causes observer's fall 
toward it, and this results in a dipole anisotropy in the CMB radiation. 
This dipole and its associated quadrupole have been subtracted 
in the results quoted so that they represent the 
true gravitational perturbation.

The anisotropy is measured by $\Delta T/\langle T\rangle$, 
where $\langle T\rangle$ is the mean temperature 
over the whole sky, and $\Delta T$ is the difference 
between the actual temperature and $\langle T\rangle$ 
for a given direction. 
In the papers, the results are given in the form of 
complete graphs, from which only a few numbers are quoted here.
For voids, the maximal difference in $\Delta T/\langle T\rangle$ 
between different directions is a few times $10^{-6}$ 
(achieved for a model with $\Omega = 0.2$, the maximal 
present density contrast $-0.75$, present radius of the 
void $ 30 h^{-1}$Mpc (similar to Bo\"otes void) 
and located at redshift between 1.25 and 8.4. 
For a condensation modelling the Virgo cluster the 
maximal difference is approx. $7\cdot 10^{-7}$ 
(almost identical for $\Omega = 0.1$ and $\Omega = 0.2)$. 
The largest maximal difference in $\Delta T/\langle T\rangle$, equal
to approx. $4.5\cdot 10^{-5}$, was found for a model 
of the Great Attractor. In this last case, such models were 
chosen which produce the peculiar velocity of the observer 
$570 \pm 60$ km/s at the radial distance $R_{0} = 43 h^{-1}$Mpc, 
while the $\Omega $-parameter of the background, 
the peculiar velocity $V$ of the observer and the 
distance to the observer were varied. 
The model that yielded the maximal value of 
$\Delta T/\langle T\rangle$ had $\Omega = 0.15$, $V = 570$ km/s 
and the distance from the observer corresponding to the redshift 8.7. 
(The values of $\Delta T/\langle T\rangle$ found by Panek for 
all three kinds of inhomogeneities, with $\Omega = 1$ 
in the background, were even lower.) 
The angular scale of the perturbation was approx. $90^{\circ }$ 
the Virgo cluster, and approx. $10^{\circ }$ 
for the Great Attractor and big voids.

The authors of the papers quoted refrained 
from drawing philosophical conclusions, but this author is tempted to 
say the following. 
Isotropy of the CMB radiation has been invoked as an 
argument in favour of the cosmological principle already 
30 years ago, when the precision of the measurement was 
at the level of $10^{-2}$. 
It has usually been said that "if the Universe were not homogeneous, 
then the inhomogeneities would leave an
imprint in the CMB radiation". 
Today, the error-bars stand at $10^{-5}$ (see Mather et al. 1992) 
and at this level they barely make contact with the expected size 
of the real effect. 
Before Panek's (1992) paper, no quantitative estimates of 
this effect had been available and all such "arguments" have been 
uncritically repeated propaganda. 
Our favourite ideologies overshadow facts also in science, 
from time to time.

\bigskip

\noindent {\bf 5. Evolution of voids}

This application is one of the most beautiful pieces of (astro-)physics done on the basis of the LT model, and the so far most conclusive results have been obtained by H. Sato and coworkers, which justifies the presence
of this contribution in this volume. 
The main references are Occhionero et al. (1978, 1981 and 1983), Maeda et al. (1983), Sato et al. (1982, 1983 and 1984){\it ,} Suto et al. (1984), with later contributions by Lake and Pim (1985), Pim and Lake (1986 and 1988), Bonnor and Chamorro (1990 and 1991) and Chamorro (1991).

A void is modelled by the LT solution as follows. It is assumed that matter is distributed homogeneously (i.e. a Friedmann model applies) around the center of symmetry out to a certain distance $r = r_{1}$. 
There it goes over into an LT region where density increases with distance. 
This region extends to $r = r_{2}$, and for $r > r_{2}$ again a Friedmann model applies, but with matter-density greater than in the central region. 
The boundaries between the three regions are not necessarily comoving with the dust. 
This configuration was used in the papers of the Occhionero and Sato groups. 
In the Lake approach the central homogeneous region contained FLRW radiation instead of dust, and in the Bonnor approach the central region was empty
(i. e. with Minkowski geometry) and all boundaries were comoving.

In great abbreviation, the results of H. Sato and coworkers are these:

1. For voids, spherical shape is stable, unlike in collapse.

2. The rim of the void propagates outward with respect to the Friedmann dust, changes into a high-density layer, and, in a finite time, because of pressure being zero, becomes a shell-crossing singularity (see next section for a definition). 
Motion of the shell can then be studied either in the
Newtonian approximation (in which case it is governed by the Sedov equation of blast waves) or by the Israel formalism in relativity. 
The singularity is not formed when there are nonzero gradients of pressure, but there exists no exact solution which could describe this process, this case was
modelled only numerically (Suto et al. 1984).

3. With a $k = 0$ background ("background" refers to the outside Friedmann region), shells smaller than the particle horizon accelerate their expansion to a finite asymptotic velocity. 
Shells larger than the particle horizon let themselves be overtaken by the horizon and then do the same.

4. With a $k < 0$ background, motion of the shell asymptotically freezes into the background flow.

5. With a $k > 0$ background, motion of the shell accelerates toward the velocity of light at recollapse. 
The void will encompass the whole space before recollapse if it was sufficiently large initially.

If a void is filled with the FLRW radiation instead of a low-density dust, then it accelerates toward the particle horizon also in the $k < 0$ background (Pim and Lake 1986). 
This implies that the observed voids must be significantly younger than the Universe.

Also here, the opposition between facts and ideology finds an illustration in the literature. 
N. R. Sen (1934 and 1997) used the LT model to investigate the stability of the closed Friedmann model against a
perturbation of matter flow which leads to a local decrease in density. 
His result, in his own words, was that "$\ldots $the [Einstein and Friedmann] models are unstable for initial rarefaction$\ldots $". 
A similar result (instability against a negative-amplitude perturbation in density) was obtained by Tolman
(1934). 
This was as close as one could get in 1934 to a prediction that voids should be expected to form in the real Universe if the Friedmann model is realistic. 
Unfortunately, the astronomical opinion-makers have known better - until voids were actually discovered in the late 1970ies.

\bigskip

\noindent {\bf 6. Singularities}

Two kinds of singularity can occur in the LT model:

1. The shell-crossing singularity, at which $R,_{r} = 0$.

2. The Big Bang/Big Crunch singularity, at which $R = 0$ while $M,_r \neq 0$.

\noindent Both are curvature singularities and $\rho \to \infty $ at both. 
However, the shell-crossing singularity is considered less dangerous. 
It results only because $p = 0$ and should go away in models in which pressure has spatial gradients (so far only numerical proofs of this conjecture are available, see Suto et al. 1984). 
It is not strong enough to be a problem for the cosmic censorship paradigm (see Joshi 1993 for a definition of strength and for a detailed discussion of cosmic censorship and related problems).
Moreover, extensions of class $C^{0}$ through this singularity exist (see Clarke and O'Donnell 1993), and in the extended spacetime the shell-crossing singularity becomes just a caustic of dust-flow.

In this section we shall deal with the Big Bang/Big Crunch singularity, which is unavoidable when $\Lambda = 0$. 
It is immediately seen, as mentioned in sec. 2, that it is in general not simultaneous in the standard cosmological synchronization. 
The simplest consequence of this property is that different regions of the Universe need not be of the same age. 
This fact was exploited by Novikov (1964) and Neeman and Tauber (1967) who tried to explain the sources of energy in quasars as "cores of delayed expansion" - i.e. the neighbourhoods where the Big Bang has occurred later than elsewhere. 
This explanation was abandoned because of discrepancies between theory and observation in various details.

The singularities in the LT model have been investigated mostly in connection with the cosmic censorship conjecture. 
The earliest counterexample to the simplest formulation of this conjecture was found by Yodzis et al (1973). 
They showed that with appropriately chosen initial conditions the shell-crossing singularity can be globally naked for a finite period of its existence. 
Eardley and Smarr (1979) then found out during numerical investigations that a globally naked singularity can be created at the center of symmetry where $R = 0 = M,_r$. 
With a certain set of initial conditions, in a certain moment the central dust particle located at $R = 0$ begins to move along a null curve on which the curvature is infinite. 
From a finite segment of this null singularity null rays can be sent to future null infinity, thus making the singularity globally naked. 
The singular null line (usually called "shell focusing singularity") later hits the spacelike surface of the Big Crunch singularity. 
In the comoving coordinates in which the LT model is most often represented the whole shell focusing singularity is mapped into a single point, making it seem to be a part of the Big Crunch singularity. 

Eardley and Smarr found this result for the $E = 0$ LT model, later many more examples of such singularities were found in the $E \neq 0$ models, see Christodoulou (1984), Newman (1986), also Krasi\'nski (1997) and Joshi (1993).

The Big Bang singularity in the LT model demonstrates one more kind of instability in the Friedmann models, first observed by Szekeres (1980), and then investigated by Hellaby and Lake (1984 and 1985b). 
In the Friedmann models, every light ray emitted from the Big Bang reaches every observer with an infinite redshift so that the Big Bang is not physically observable. 
In the LT model this is true for those locations at the Big Bang where $t_{B,r} \to 0$ sufficiently rapidly, and also for nonradial rays and noncomoving emitters. 
Radial rays emitted from comoving sources reach every observer with an infinite blueshift.

The equation $R,_{r} = 0$ is a necessary condition for a shell-crossing singularity, but not a sufficient one. 
If the set $\{t, r_{S}(t))$ on which $R,_{r} = 0$ has in addition the following properties:

\begin {equation}
r_{S,t} = 0 = M,_{r}(r_{S}) = t_{B,r}(r_{S}) = E,_{r}(r_{S}),\qquad E(r_{S}) = 1/2, 
\end {equation}

\noindent then the hypersurface $R,_{r} = 0$ (which then consists of the flow lines of dust) is nonsingular, and it is a locus of extrema of the function $R(t,r)$. 
This locus is called a neck. 
The existence of necks was first pointed out by Hellaby and Lake (1985a and c), and several unusual properties offered by them have been discussed by Hellaby (1987). 
One of these is the existence of a "string of beads" Universe which begins as a chain of isolated white holes that keep growing until they come into contact through necks, then evolve for a finite time, separate to form
isolated black holes and collapse to their separate Big Crunches.

As should be seen from this very short overview, also for singularities the FLRW models present an oversimplified picture and do not offer any insight into the more general cases.

\bigskip

\noindent {\bf 7. Influence of electromagnetic field on the collapse of dust}

Properties of charged dust and of neutral dust moving in an exterior electromagnetic field (in both cases under the assumption of spherical symmetry) have been extensively discussed in several papers, the most important references are Markov and Frolov (1970) and Vickers (1973). 
In the comoving coordinates the Einstein-Maxwell equations form a coupled set that is difficult to handle (the set can be solved explicitly in noncomoving coordinates, Ori 1990). 
Still, it was shown that if $\Lambda = 0$ and the charge density on the dust particles is {\it smaller} than the mass-density (in units in which $c = G = 1)$, then the Big Bang/Big Crunch singularity will not occur. 
(Large charge density enhances gravitation by its
contribution to the total energy-density.) 
The Big Bang/Crunch is avoided also for neutral dust with $\Lambda = 0,$ if it moves in an
exterior spherically symmetric electromagnetic field, created by a charge placed at the center or charges distributed in a finite volume around the center (Shikin 1972). 
In this last case the metric is given by eqs. (1) and (2), but the second of eqs. (2) is replaced by:

\begin{equation}
{R,_t}^2 = 2E(r) + 2M(r)/R - Q^{2}/R^{2} + {1\over 3} \Lambda R^{2},
\end{equation}

\noindent where $Q$ is the (constant) value of the electric charge. Hence, the electromagnetic field generates a repulsive gravitational field.

This result created an expectation that a finite volume of charged dust could tunnel through the wormhole in the maximally extended Reissner - Nordstr\"om spacetime to another asymptotically flat sheet. 
This expectation was killed by the papers of Ori (1990 and 1991) who showed that while the dust is already under the inner horizon of the Reissner - Nordstr\"om spacetime, shell-crossings necessarily occur that block the passage through the wormhole. 
This happens because the dust layers that have collapsed to a smaller radius bounce first and collide with the next layers that still go on collapsing. 
The shell-crossings might possibly be avoided with nonzero pressure-gradients, but no such exact solutions are available so far.

\bigskip

\noindent {\bf 8. Other properties}

Several interesting properties of the LT model have been omitted in this short overview (see Krasi\'nski 1997 - that one was meant to be a complete review). 
Among the more important ones are:

1. The creation and evolution of a black hole can be studied (Barnes 1970). 
In the Schwarzschild and Kerr solutions, the black hole is stationary and present in unchanged form throughout the whole history of the spacetime. 
This does not allow for a description of the black hole's origin and of the process of accretion of matter onto it. 
In the LT model, one can assume such initial conditions that the black hole does not exist initially, then appears and keeps growing. 

2. If it is assumed that the cosmic medium (the average matter density in the Universe) extends throughout the solar planetary system, then circular orbits of the planets are not possible (Gautreau 1984). 
All orbits must expand by the law (calculated in the Newtonian approximation):

\begin{equation}
{d{\cal R}\over dt} = 8\pi {\cal R}^{4}H\overline{\rho}/(2M),
\end{equation}

\noindent where $H$ is the Hubble parameter, $\overline{\rho}$ is the average cosmic mass-density, $M$ is the mass of the Sun and ${\cal R}$ is the radius of the orbit. 
For Saturn, this rate of expansion is $6\cdot 10^{-18}$ meters/year - unmeasureably small, but nonzero.

It is commonly believed that the last word of the relativity theory on this subject were the two papers by Einstein and Straus (1945 and 1946), in which a planetary system was assumed to be a portion of the Schwarzschild spacetime matched to a Friedmann background. 
The result was that the expansion of the Universe does not influence the planetary orbits. 
The Gautreau paper shows that this is not the only possible model. 
Moreover, the papers by Sato et al. mentioned in sec. 5 demonstrate that the configuration used by Einstein and Straus is unstable. 
The matching condition is a unique connection between the Schwarzschild mass and the mass removed from the Friedmann model to make place for the Schwarzschild region. 
For the $k = 0$ Friedmann background the two masses have to be simply equal. 
The LT model allows one to consider the uncompensated situation, when the mass in the vacuole is initially larger or smaller than the Einstein-Straus value. 
This was done in some of the papers on evolution of voids - and it turned out that when the Einstein-Straus matching condition is not fulfilled, the edge of the void is not comoving with the Friedmann dust. 
It will either expand faster than the Friedmann background (when the mass in the void is "too small") or collapse toward its center.

3. Galaxies could not have formed out of statistical fluctuations in an initially homogeneous matter-density because the Universe is much too young to have allowed enough time for this (Bonnor 1956). 
In order to develop into galaxies today, the initial perturbations would have to be $10^{29}$ times larger than the statistical fluctuations. 
This inference is usually associated with the perturbative approach to Einstein's theory which had not yet really existed before 1956, while Bonnor came to this conclusion on the basis of an exact solution.

\bigskip

\noindent {\bf 9. Conclusion}

The LT model was discussed here because it is the most mature one from the point of view of astrophysics among the several generalizations of the FLRW models that exist in the literature. 
However, many other generalizations have been found, and many papers were published on their properties. 
(The main review, Krasi\'nski 1997, contains more than 700 references. Readers are referred there for the literature on the subjects mentioned below.) 
In particular, the plane- and hyperbolically symmetric counterparts of the LT model are known, also with a nonzero cosmological constant. 
The same is true for the Datt (1938) model. 
All these models were generalized for electromagnetic fields of the respective symmetry. 
Generalizations with viscosity and heat-flow are also known. 
Szekeres (1975) found a generalization of the $\Lambda = 0$ LT and Datt models that has no symmetry. 
The Szekeres solutions were later generalized for nonzero $\Lambda$ and nonzero pressure (which must have its gradient collinear with the fluid velocity).

The solutions mentioned above are characterized by zero acceleration and zero rotation of the fluid source. 
An even larger family of generalizations of the FLRW models exists in which shear and rotation are zero. 
One subclass of them is conformally flat, these are all known explicitly. 
In the conformally nonflat subclass, the Einstein equations were reduced to a single ordinary differential equation of second order; a large body of literature exists on particular solutions of this equation. 
Also in this family, generalizations with electromagnetic field and heat-flow are known. 

Large families of generalizations of the FLRW models are known in which 1. The Vaidya radiation is present along with a perfect fluid; 2. The source is a "stiff fluid" with a two-dimensional commutative symmetry group; 3. The spacetime is a generalization of the Senovilla singularity-free solution. This list does not include isolated experiments with finding exact solutions that cannot be organized into correlated families.

The overall conclusion from the present article should be that investigation of inhomogeneous models of the Universe is already a highly developed branch of relativity. 
It is time to begin taking all this knowledge seriously. 
Pretending that the FLRW models is everything that relativity has to say on cosmology is counterproductive and contradicts facts. 
This attitude is particularly harmful when it is spread through newly published textbooks.

\bigskip

\noindent {\bf Acknowledgements}

I am grateful to A. Barnes, W. B. Bonnor, C. J. S. Clarke, D. Eardley, G. F. R. Ellis, C. Hellaby, H. Kurki-Suonio, K. Lake, J. Moffat, D. S\'aez, H. Sato and P. Szekeres for the permissions to use their figures in my talk at the Yamada conference, 
and to D. S\'aez for very helpful comments on the present text.

\bigskip

\noindent {\bf References}

\noindent Arnau, J. V., Fullana, M., Monreal, L., S\'aez, D. (1993), {\it Astrophys. J.} {\bf 402}, 359.
 
\noindent Arnau, J. V., Fullana, M., S\'aez, D. (1994), {\it Mon. Not. Roy. Astr. Soc.} {\bf 268}, L17.
 
\noindent Barnes, A. (1970), {\it J. Phys}. {\bf A3}, 653.
 
\noindent Bonnor, W. B. (1956), {\it Z. Astrophysik} {\bf 39}, 143.
 
\noindent Bonnor, W. B., Chamorro, A. (1990), {\it Astrophys. J.} {\bf 361}, 21.
 
\noindent Bonnor, W. B., Chamorro, A. (1991), {\it Astrophys. J.} {\bf 378}, 461.
 
\noindent Broadhurst, T. J., Ellis, R. S., Koo, D. C. and Szalay, A. S. (1990), {\it Nature} {\bf 343}, 726.
 
\noindent Chamorro, A. (1991), {\it Astrophys. J.} {\bf 383}, 51.
 
\noindent Christodoulou, D. (1984), {\it Commun. Math. Phys.} {\bf 93}, 171.

\noindent
Clarke, C. J. S., O'Donnell, N. (1992), {\it Rendiconti del Seminario Matematico della Universita e Politecnico de Torino} {\bf 50}(1), 39.

\noindent
Datt, B. (1938), {\it Z}. {\it Physik} {\bf 108}, 314.

\noindent
Eardley, D. M., Smarr, L. (1979), {\it Phys. Rev.} {\bf D19}, 2239.

\noindent
Einstein, A., and Straus, E. G. (1945), {\it Rev. Mod. Phys.} {\bf 17}, 120.

\noindent
Einstein, A., and Straus, E. G. (1946), {\it Rev. Mod. Phys.} {\bf 18}, 148.

\noindent
Ellis, G. F. R. (1984), in: {\it General relativity and gravitation}. Edited by B. Bertotii, F. de Felice, A. Pascolini. D. Reidel, Dordrecht, p. 215.

\noindent
Fullana, M. J., Arnau, J. V., and S\'aez, D. (1996), {\it Mon. Not. Roy. Astr. Soc.} {\bf 280}, 1181.

\noindent
Gautreau, R. (1984), {\it Phys. Rev.} {\bf D29}, 198.

\noindent
Hellaby, C. (1987), {\it Class. Q. Grav.} {\bf 4}, 635.

\noindent
Hellaby, C., Lake, K. (1984), {\it Astrophys. J.} {\bf 282}, 1.

\noindent
Hellaby, C., Lake, K. (1985a), {\it Astrophys. J.} {\bf 290}, 381.

\noindent
Hellaby, C., Lake, K. (1985b), {\it Astrophys. J.} {\bf 294}, 702.

\noindent
Hellaby, C., Lake, K. (1985c), {\it Astrophys. J.} {\bf 300}, 461.

\noindent
Joshi, P. S. (1993), {\it Global aspects in gravitation and cosmology.} Clarendon Press, Oxford, pp. 242 - 255.

\noindent
Kantowski, R., Sachs, R. K. (1966), {\it J. Math. Phys.} {\bf 7}, 443.

\noindent
Krasi\'nski, A. (1997), {\it Inhomogeneous cosmological models.} Cambridge University Press.

\noindent
Kurki-Suonio, H., Liang, E. (1992), {\it Astrophys. J.} {\bf 390}, 5.

\noindent
Lake, K., Pim, R. (1985), {\it Astrophys. J.} {\bf 298}, 439.

\noindent
Lema\^{i}tre, G. (1933), {\it Ann. Soc. Sci. Bruxelles} {\bf A53,} 51. [English translation: Lema\^{i}tre, G. (1997), {\it Gen. Rel. Grav.} {\bf 29,} 641].

\noindent
Maeda, K., Sasaki, M., Sato, H. (1983), {\it Progr. Theor. Phys.} {\bf 69}, 89.

\noindent
Maeda, K., Sato, H. (1983a), {\it Progr. Theor. Phys.} {\bf 70}, 772.

\noindent
Maeda, K., Sato, H. (1983b), {\it Progr. Theor. Phys.} {\bf 70}, 1276.

\noindent
Markov, M. A., Frolov, V. P. (1970), {\it Teor. Mat. Fiz}. {\bf 3}, 3 [{\it Theor. Math. Phys.} {\bf 3}, 301 (1970)].

\noindent
Mather, J. C., Bennett, C. L., Boggess, N. W., Hauser, M. G., Smoot, G. F., Wright, E. L. (1993), in: {\it General Relativity and Gravitation 1992.} Proceedings of the 13th International Conference on General Relativity and Gravitation at Cordoba 1992. Edited by R. J. Gleiser, C. N. Kozameh and O. M. Moreschi. Institute of Physics Publishing, Boston and Philadelphia, p. 151.

\noindent
Moffat, J. W., Tatarski, D. C. (1992), {\it Phys. Rev.} {\bf D45}, 3512.

\noindent
Moffat, J. W., Tatarski, D. C. (1995), {\it Astrophys. J.} {\bf 453}, 17.

\noindent
Neeman, Y., Tauber, G. (1967), {\it Astrophys. J.} {\bf 150}, 755.

\noindent
Novikov, I. D. (1964), {\it Astron. Zh.} {\bf 41}, 1075 [{\it Sov. Astr. A. J.} {\bf 8}, 857 (1965)].

\noindent
Occhionero, F., Santangelo, P., Vittorio, N. (1983), {\it Astron. Astrophys.} {\bf 117}, 365.

\noindent
Occhionero, F., Vecchia-Scavalli, L., Vittorio, N. (1981a), {\it Astron. Astrophys.} {\bf 97}, 169.

\noindent
Occhionero, F., Vecchia-Scavalli, L., Vittorio, N. (1981b), {\it Astron. Astrophys.} {\bf 99}, L12.

\noindent
Occhionero, F., Vignato, A., Vittorio, N. (1978), {\it Astron. Astrophys.} {\bf 70}, 265.

\noindent
Ori, A. (1990), {\it Class. Q. Grav.} {\bf 7}, 985.

\noindent
Ori, A. (1991), {\it Phys. Rev.} {\bf D44}, 2278.

\noindent
Newman, R. P. A. C. (1986a), {\it Class. Q. Grav.} {\bf 3}, 527.

\noindent
Newman, R. P. A. C. (1986b), in: {\it Topological properties and global structure of spacetime}. Edited by P. G. Bergmann and V. de Sabbata. Plenum, New York, p. 153.

\noindent
Panek, M. (1992), {\it Astrophys. J.} {\bf 388}, 225.

\noindent
Pim, R., Lake, K. (1986), {\it Astrophys. J.} {\bf 304}, 75.

\noindent
Pim, R., Lake, K. (1988), {\it Astrophys. J.} {\bf 330}, 625.

\noindent
Raine, D. J., Thomas, E. G. (1981), {\it Mon. Not. Roy. Astr. Soc.} {\bf 195}, 649.

\noindent
Ribeiro, M. B. (1992a), {\it Astrophys. J.} {\bf 388}, 1.

\noindent
Ribeiro, M. B. (1992b), {\it Astrophys. J.} {\bf 395}, 29.

\noindent
Ribeiro, M. B. (1993), {\it Astrophys. J.} {\bf 415}, 469.

\noindent
S\'aez, D., Arnau, J. V., Fullana, M. J. (1993), {\it Mon. Not. Roy. Astr. Soc.} {\bf 263}, 681.

\noindent
Sato, H. (1982), {\it Progr. Theor. Phys.} {\bf 68}, 236.

\noindent
Sato, H. (1984), in: {\it General Relativity and Gravitation.} Edited by B. Bertotti, F. de Felice, A. Pascolini. D. Reidel, Dordrecht, p. 289.

\noindent
Sato, H., Maeda, K. (1983), {\it Progr. Theor. Phys.} {\bf 70}, 119{\it .}

\noindent
Sen, N. R. (1934), {\it Z. Astrophysik} {\bf 9}, 215 [see reprint: Sen, N. R. (1997), {\it Gen. Rel. Grav.} {\bf 29}, 1477].

\noindent
Shikin, I. S. (1972), {\it Commun. Math. Phys.} {\bf 26}, 24.

\noindent
Suto, Y., Sato, K., Sato, H. (1984a), {\it Progr. Theor. Phys.} {\bf 71}, 938.

\noindent
Suto, Y., Sato, K., Sato, H. (1984b), {\it Progr. Theor. Phys.} {\bf 72}, 1137.

\noindent
Szekeres, P. (1975), {\it Commun. Math. Phys.} {\bf 41}, 55.

\noindent
Szekeres, P. (1980), in: {\it Gravitational radiation, collapsed objects and exact solutions}. Edited by C. Edwards. Springer (Lecture notes in physics, vol. 124), New York, p. 477.

\noindent
Tolman, R. C. (1934), {\it Proc. Nat. Acad. Sci. USA} {\bf 20}, 169. [See reprint: Tolman, R. C. (1997), {\it Gen. Rel. Grav.} {\bf 29}, 935].

\noindent
Vickers, P. A. (1973), {\it Ann. Inst. Poincare} {\bf A18}, 137.

\noindent
Yodzis, P., Seifert, H. J., M\"uller zum Hagen, H. (1973), {\it Commun. Math. Phys.} {\bf 34}, 135.

\end {document}